\documentclass[12pt]{article}
\pdfoutput=1

\usepackage{jheppub}
\makeatletter
\def\@fpheader{\relax}
\makeatother

\usepackage{caption}
\usepackage{subcaption}
\usepackage{amsmath,amssymb,amsfonts}
\usepackage{graphicx}
\usepackage{multicol}
\usepackage{multirow}
\usepackage{float}
\usepackage[table]{xcolor}

\newcommand\be{\begin{equation}}
\newcommand\ee{\end{equation}}
\newcommand\bea{\begin{eqnarray}}
\newcommand\eea{\end{eqnarray}}
\newcommand\ba{\begin{array}}
\newcommand\ea{\end{array}}
\newcommand\nn{\nonumber}
\newcommand\bra[1]{\langle#1|}
\newcommand\ket[1]{|#1\rangle}
\newcommand\comment[1]{}
\newcommand\NatureComment[1]{#1}
\newcommand\sctn[1]{\section{#1}}
\newcommand\subsec[1]{\section{#1}}
\newcommand\cttn[1]{~\cite{#1}}

\title{Baryons from Mesons: A Machine Learning Perspective}

\author[a,b]{Yarin Gal}
\author[c]{\!\!, Vishnu Jejjala}
\author[c,d]{\!\!, Dami\'an Kaloni Mayorga Pe\~na}
\author[a,b]{\!\!, Challenger Mishra}

\affiliation[\,a]{OATML, Department of Computer Science, University of Oxford, Oxford OX1 3QD, United Kingdom}
\affiliation[\,b]{The Alan Turing Institute, British Library, London NW1 2DB, United Kingdom}
\affiliation[\,c]{Mandelstam Institute for Theoretical Physics, School of Physics, NITheP, and CoE-MaSS, \\ University of the Witwatersrand, Johannesburg, WITS 2050, South Africa}
\affiliation[\,d]{Data Laboratory, Universidad de Guanajuato,
Loma del Bosque No. 103 Col. Lomas del Campestre C.P 37150 Leon, Guanajuato, Mexico}

\emailAdd{yarin.gal@cs.ox.ac.uk}
\emailAdd{vishnu@neo.phys.wits.ac.za}
\emailAdd{damian.mayorgapena@wits.ac.za}
\emailAdd{challenger.mishra@gmail.com}

\abstract{
Quantum chromodynamics (QCD) is the theory of the strong interaction.
The fundamental particles of QCD, quarks and gluons, carry colour charge and form colourless bound states at low energies.
The hadronic bound states of primary interest to us are the mesons and the baryons.
From knowledge of the meson spectrum, we use neural networks and Gaussian processes to predict the masses of baryons with $90.3$\% and $96.6$\% accuracy, respectively. 
These results compare favourably to the constituent quark model.
We as well predict the masses of pentaquarks and other exotic hadrons.
}

\date{}

\newcommand\eref[1]{(\ref{#1})}

\begin{document} 

\parskip=.5\baselineskip

\maketitle

\sctn{Introduction}\label{sec:intro}
In QCD, massive fermionic particles (quarks and antiquarks) interact via the exchange of massless bosons (gluons).
These particles are the local excitations of a non-Abelian quantum field theory with gauge group $SU(3)$.
The particles exhibit colour, and there are three colours in Nature ($N = 3$).
Each generation of the Standard Model of particle physics includes two quarks, and there are six quarks in total.
(The number of flavours is $N_f = 6$.)
QCD is famously asymptotically free\cttn{Gross:1973ju,Gross:1973id,Politzer:1973fx,tHooft:1998qmr}.
At energies below $\Lambda_\mathrm{QCD}\approx (332\pm17)$ MeV\cttn{PhysRevD.98.030001}, the scale at which the strong coupling $g_s$ diverges perturbatively, the theory confines, meaning that quarks and gluons combine to form colour neutral bound states, the hadrons.

Among the hadrons, mesons comprise a quark--antiquark pair while baryons consist either of three quarks or three antiquarks.
This is because quarks transform in the fundamental representation (the $\mathbf{3}$ of $SU(3)$) whereas antiquarks transform in the anti-fundamental representation (the $\overline{\mathbf{3}}$ of $SU(3)$).
The representation theory of $SU(3)$ teaches us that
\bea
\mathbf{3} \otimes \overline{\mathbf{3}} &=& \mathbf{1} \oplus \mathbf{8} ~, \nn \\
\mathbf{3} \otimes \mathbf{3} \otimes \mathbf{3} &=& \mathbf{1} \oplus \mathbf{8} \oplus \mathbf{8} \oplus \mathbf{10} ~, \label{eq:repth} \\
\overline{\mathbf{3}} \otimes \overline{\mathbf{3}} \otimes \overline{\mathbf{3}} &=& \mathbf{1} \oplus \mathbf{8} \oplus \mathbf{8} \oplus \overline{\mathbf{10}} ~. \nn
\eea
A quark in the $\mathbf{3}$ representation combines with an antiquark in the $\overline{\mathbf{3}}$ representation to form a colour singlet bound state, denoted by $\mathbf{1}$ in the direct sums on the right hand side of~\eref{eq:repth}.
The colour indices are summed over.
Similarly, three particles in the $\mathbf{3}$ representation or three particles in the $\overline{\mathbf{3}}$ also combine to form colour singlets, where the colour indices are contracted with a totally antisymmetric epsilon symbol.
\comment{
Explicitly, the mesons, baryons, and anti-baryons are, respectively,
\be
q^i_a \bar{q}_{i,b} ~, \qquad \epsilon_{ijk} q^i_a q^j_b q^k_c ~, \qquad \epsilon^{ijk} \bar{q}_{i,a} \bar{q}_{j,b} \bar{q}_{k,c} ~,
\ee
where $q$ and $\bar{q}$ are quark and antiquark fields and $i,j,k$ denote colour indices and $a,b,c$ denote flavour indices.
}

Generalising the $SU(3)$ theory to $SU(N)$ and taking $N$ large, Witten showed that baryons are solitons in the meson spectrum\cttn{Witten:1979kh}.
In the limit where $N\to \infty$, the mesons are effectively free particles with couplings of ${\cal O}(\frac{1}{N})$.
The weakly coupled effective field theory of mesons has monopole like excitations, which are identified with baryons.
As before, the mesons are a quark--antiquark bound state, but now the baryons consist of $N$ distinct quarks or $N$ distinct antiquarks ($N_f \ge N$).
Baryon masses are of ${\cal O}(N)$ in the large-$N$ expansion.
This insight motivates the baryon vertex\cttn{Witten:1998xy} in the AdS/CFT correspondence\cttn{Maldacena:1997re,Gubser:1998bc,Witten:1998qj}.

Na\"{\i}vely, $N = 3$ does not present an environment in which large-$N$ dynamics should manifest.
However, errors are typically suppressed by inverse powers of the number of colours, $N$.
In this work, we use machine learning to train on mesonic data and predict baryonic masses in $SU(3)$ QCD.
We also predict the masses of tetraquarks, pentaquarks, and other exotic states.

Machine learning has already been successful at reproducing phenomenological features of QCD with three colours.
For example, Reference~\cite{Hashimoto:2018bnb} trains on lattice QCD data for the chiral condensate vacuum expectation value vs.\ quark mass, obtains a metric from parameters in a neural network, and predicts the form of the quark--antiquark potential as a function of the separation of the constituent particles by applying the AdS/QCD dictionary\cttn{Maldacena:1998im,Rey:1998bq}.
Indeed, Reference~\cite{Hashimoto:2019bih} suggests that AdS/CFT correspondence itself recapitulates the structure of a restricted Boltzmann machine.
Given that ${\cal N}=4$ super-Yang--Mills theory, the conformal field theory dual to string theory on the AdS$_5\times S^5$ spacetime, is a toy model for QCD\cttn{Witten:1998zw}, albeit one without a mass gap, it is reasonable to apply machine learning techniques to investigate the low energy spectrum of the strong interaction.

For the computation of hadronic masses, particle physicists appeal to a variety of models ranging from the constituent quark model to sophisticated lattice QCD computations.
Here, we compare our results to the simplest constituent quark model in which QCD effects are included in the form of effective quark masses leading to reasonable agreement with the baryon spectrum.
Of course, with lattice QCD, one expects to get better approximations to hadron masses but usually at the expense of involved computations.
In this work, we propose machine learning methods as a \emph{via di mezzo}.
Considering only the quark composition and global attributes of the particle, we are able to obtain results significantly better than a simple linear regression.

\NatureComment{
The organisation of this paper is as follows.
In Section~\ref{sec:two}, we discuss our dataset of mesons and baryons.
In Section~\ref{sec:three}, we describe the machine learning techniques and experimental set-up. 
In Section~\ref{sec:four}, we present the results of our analysis.
Finally, in Section~\ref{sec:five}, we discuss our results.
}

\subsec{Mesons and baryons}\label{sec:two}
Hadrons are bound states of quarks and gluons.
We identify these by an input $13$-vector
\be
\vec{v} = (d,\bar{d},u,\bar{u},s,\bar{s},c,\bar{c},b,\bar{b},I,J,P) ~. \label{eq:vec}
\ee
The first ten entries in the vector count the number of valence quarks of a given type; $I$ is isospin, $J$ is angular momentum, and $P$ is parity.
Isospin and angular momentum are positive half-integers.\footnote{From a purely machine learning perspective, it is worth stating the range of values for the components of $\vec{v}$. In the particle datasets we consider in this paper, $v_{1\le i\le 10}\in\{0,1,2,3,4\}$, $I\in \{0,\frac{1}{2},1,\frac{3}{2}\}$, $J \in \{0,\frac{1}{2},1,\frac{3}{2},\dots,4\}$, and $P\in\{-1,1\}$.}
Because of the incompleteness of measurements, we do not include $G$-parity, $C$-parity, or width.
There are no known hadrons with $t$ or $\bar{t}$ valence quarks.
The charge of the particle is the sum of the charges of the valence quarks ($+\frac23$ for up and charm and $-\frac13$ for down, strange, and bottom), and is therefore redundant information that is not included.
The charges of the antiquarks, denoted with a bar, are the opposite of the charges of the quarks.

We thus encode the charged pions, $\pi^+ = u\bar{d}$ and $\pi^- = \bar{u}d$, as
\bea
\pi^+ = (0,1,1,\vec{0}_7,1,0,-1) ~, \qquad &&
\pi^- = (1,\vec{0}_2,1,\vec{0}_6,1,0,-1) ~,
\eea
where $\vec{0}_k$ is a zero vector of length $k$.
For mesons that are linear combinations of pairs of quarks and antiquarks, we adhere to the numbering scheme in the Particle Data Book\cttn{PhysRevD.98.030001} to specify the valence quarks.
Of note, according to these conventions, the quark numbers of the isotriplet mesons ($\pi^0$, $\rho^0$, $\ldots$) are $(1,1,\vec{0}_8,I,J,P)$, the quark numbers for the lighter isosinglet mesons ($\eta$, $\omega$, $\ldots$) are $(\vec{0}_2,1,1,\vec{0}_6,I,J,P)$, and the quark numbers for the heavier isosinglet mesons ($\eta'$, $\phi$, $\ldots$) are $(\vec{0}_4,1,1,\vec{0}_4,I,J,P)$.
In total, there are $196$ mesons in the dataset.
Particles and antiparticles such as $\pi^+$ and $\pi^-$ are counted separately in this enumeration.
We do not \textit{ab initio} insist that particles and their antiparticles have the same mass; this is a fact that we leave to the machine to learn.

The $43$ baryons in the dataset are similarly encoded.
The light baryons of particular interest to us are the proton and the neutron:
\bea
p = (1,0,2,\vec{0}_7,\frac12,\frac12,1) ~, \qquad &&
n = (2,0,1,\vec{0}_7,\frac12,\frac12,1) ~. 
\eea

\subsec{Machine learning hadrons}\label{sec:three}
We employ two basic tools for predicting baryon masses from mesonic data --- \textbf{neural networks} and \textbf{Gaussian processes} --- which we briefly summarise in this section. Following this, we detail the set-up of our machine learning experiments. 

\subsection{Neural networks}
Our neural network associates an input vector~\eref{eq:vec} to the mass of the particle.
To construct the first hidden layer of a feed-forward neural network, we define
\be
\vec{a}^{(1)} = W^{(1)} \cdot \vec{v} + \vec{b}^{(1)} ~,
\ee
where $W^{(1)}$ is an $n_1\times 13$ matrix $W^{(1)}$ and $b^{(1)}$ is an $n_1$-vector.
We then act on the components of $\vec{a}^{(1)}$ elementwise with a non-linear activation function $\sigma^{(1)}$.
The next hidden layer of the neural network starts from this new vector and computes
\be
\vec{a}^{(2)} = W^{(2)} \cdot \sigma^{(1)}(\vec{a}^{(1)}) + \vec{b}^{(2)} ~,
\ee
where $W^{(2)}$ is an $n_2\times n_1$ matrix and $ \vec{b}^{(2)}$ is an $n_2$-vector;
we then apply a new activation function $\sigma^{(2)}$ elementwise to the components of $\vec{a}^{(2)}$.
This process iterates over $\ell$ hidden layers, and we take the sum of the entries of the final $n_\ell$-vector to give a number, $f_\theta(\vec{v})$.
Here, $\theta$ is a collective label that denotes the elements of the weight matrices $W^{(i)}$ and bias vectors $\vec{b}^{(i)}$.

We fit the parameters, $\theta$, of the neural network using a loss function with respect to a training set $S$, which contains relations between inputs and outputs, $\vec{v}_s \rightarrow \log m_s$. A typical loss function is the \emph{mean squared loss} defined as
\be
g(\theta) = \sum_{s} || f_\theta(\vec{v}_s) - \log m_s ||^2 ~.
\label{eq:gth}
\ee
In our machine learning experiments with neural networks (which we detail in \ref{exptsetup}), the set $S$ always corresponds to the mesonic data, or a subset of it. As such, $\vec{v}_s$ and $\log m_s$ stand for the input encoding of the mesons and their (logarithmic) masses respectively. 

Neural networks tend to overfit quickly: they predict well on the training set but badly on the test set. This effect is especially pronounced with small training sets such as ours. Therefore we use a validation set ($S^c$) disjoint from the training set ($S$) and test set (baryons) for tuning hyperparameters, optimisation, and selecting model architecture from a class of neural networks. This is done as follows. First, the entire mesonic data is split into $k$ (approximately) equal parts, $(k-1)$ of which constitute \emph{training} data and the remaining, \emph{validation} data. The network loss (for each member of the class of neural networks) is then minimised with respect to the training data (via backpropagation) and the error computed on the validation data. This is repeated $k$ times corresponding to the $k$ choices for the training/validation split, and the error is averaged over. The network with the smallest average validation error is then chosen. This network is then trained using the full mesonic data and tested on the baryonic data. 

\subsection{Gaussian processes}
\begin{figure}[t!]
\centering
\includegraphics[scale=.4]{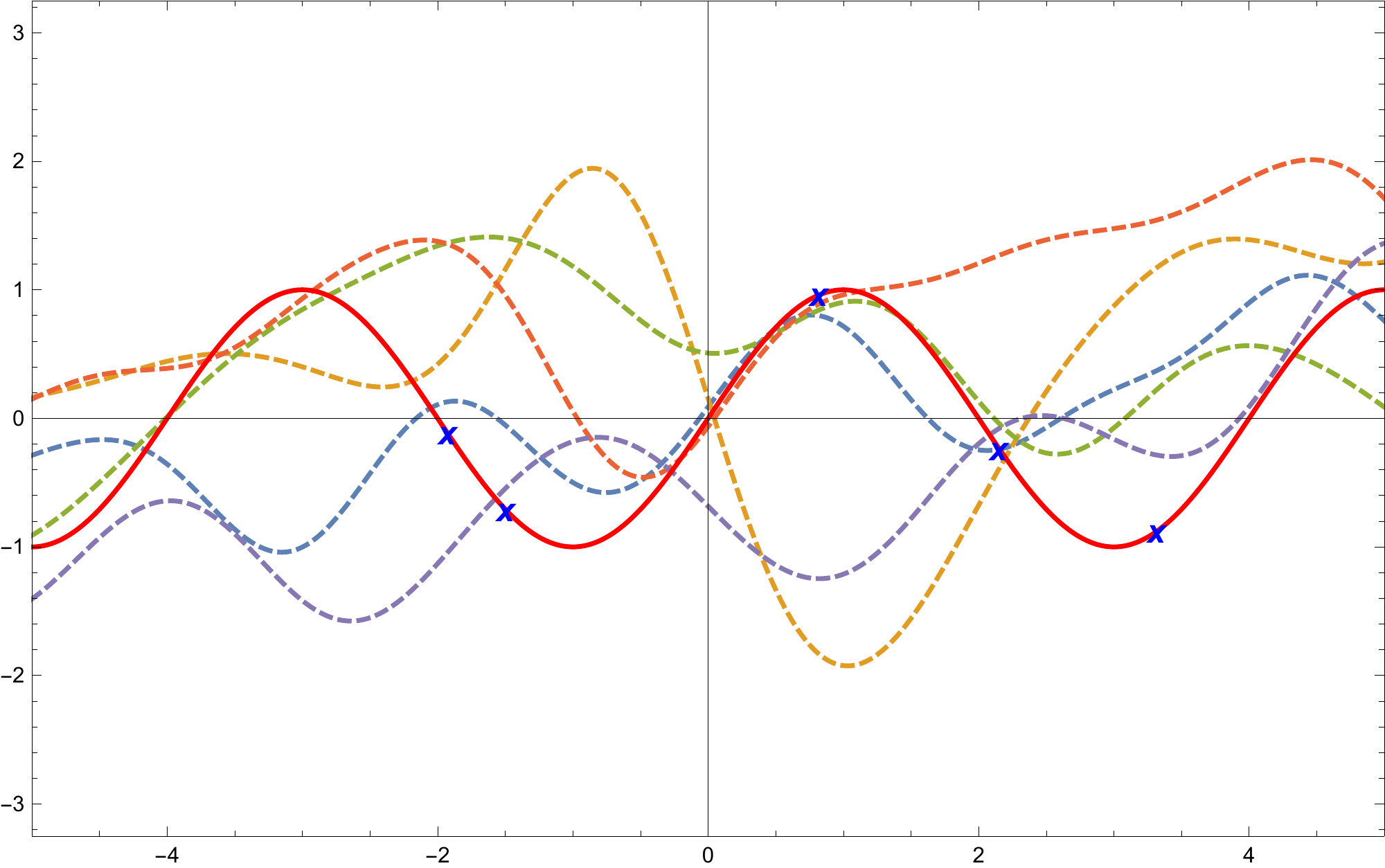}\\[20pt]
 \includegraphics[scale=.4]{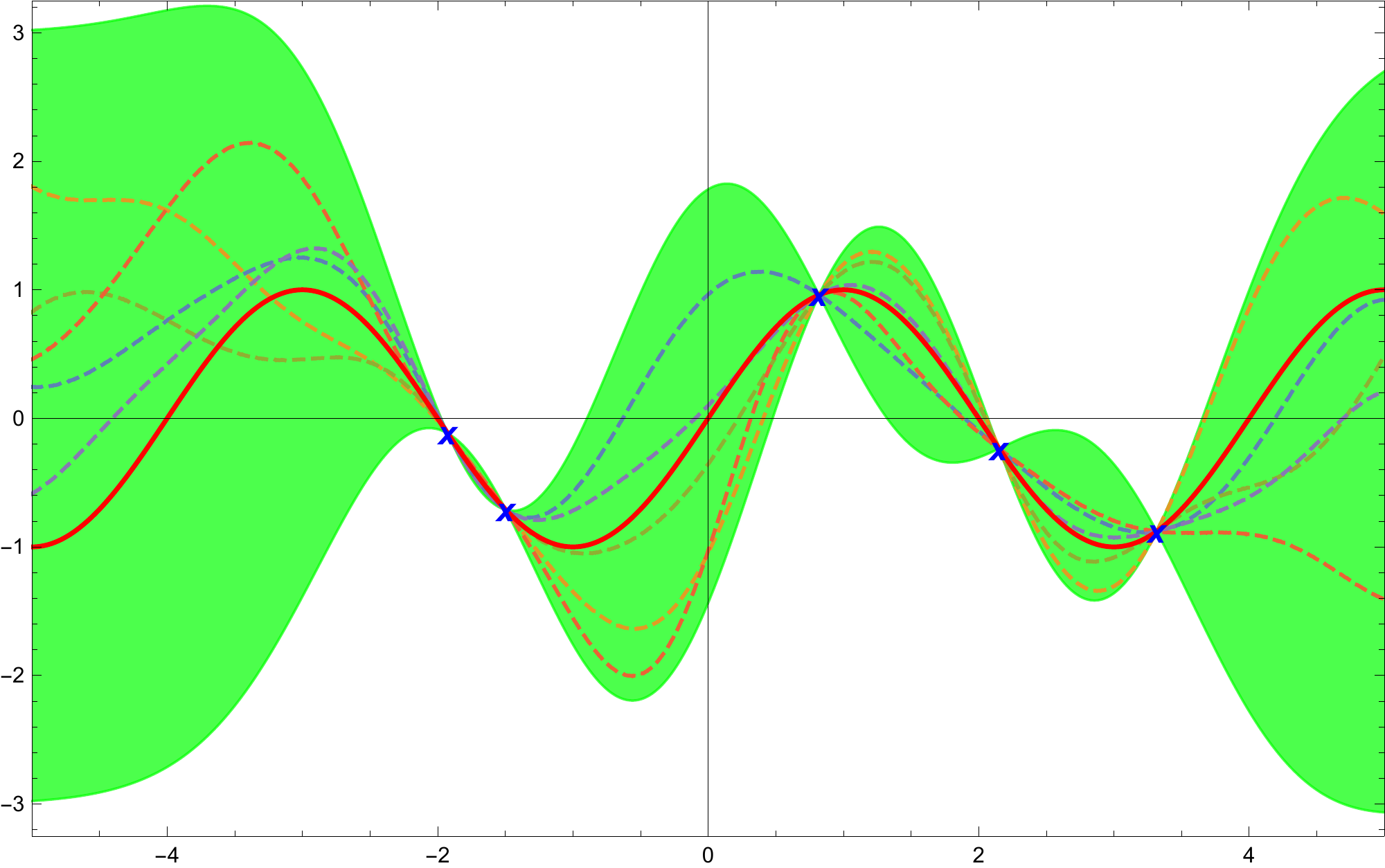}\\
\caption{\small\textit{An illustration of inference using Gaussian processes. 
A squared exponential kernel is used to fit the function ${\sin \left(\frac{\pi}{2}x\right)}$ (shown by the solid red curves) with function values provided at five points (shown in blue). In the top panel, the dashed curves are functions drawn from the Gaussian process prior with zero mean.  In the bottom panel, the dashed curves were sampled from the Gaussian process posterior. The green region in the bottom  panel denotes the $99$\% confidence intervals for the predictions. The variance in predictions increases as one moves further away from points where function values are known.} }
  \label{fig:GPtoy}
  \vspace{-10pt}
\end{figure}

Gaussian processes provide a powerful tool for making statistical inference.
While we refer the reader to Reference\cttn{gpml} for more details, briefly,
a Gaussian process is a collection of random variables such that any subset of them is normally distributed.
The Gaussian process is fully specified by a covariance and a mean function.
When used as a tool in regression problems, the output of a Gaussian process is a normal distribution instead of a point estimate, a key difference from a non-Bayesian neural network.
In particular, a Gaussian process supplies the mean and the variance of the normal distribution, which can then be used to estimate confidence in predictions.
Instead of outputting a specific function, the Gaussian process posterior provides a distribution from which functions can be drawn, each of which will fit the training data to high accuracy. 
When applied to unseen data, the variance of a Gaussian process prediction will be small if the data is close to the training data.
Conversely, the variance is large when the data is far away from the training data.
This is illustrated in Figure~\ref{fig:GPtoy}.
The Gaussian process is equivalent to a fully connected single layer neural network with an independent and identically distributed (i.i.d.)\ prior over its parameters in the limit of infinite width\cttn{Neal}.
Thus, Gaussian processes could be used to make exact Bayesian inferences for infinite width neural networks. 

Making inference using a Gaussian process involves inverting matrices that are the size of the dataset. A Gaussian process is therefore particularly well-suited to relatively small datasets, such as ours. Further, unlike neural networks, Gaussian processes do not require a validation set to tune hyperparameters. Instead, Gaussian processes use their own \emph{marginal likelihood} --- a quantity that captures how well each subset of the training points can predict the rest of the training points.

Constructing a Gaussian process proceeds as follows. We first specify a positive definite \emph{kernel} 
\be\label{eq:Gaussian processKernel}
K_{ij} := k(x_i, x_j)~,
\ee
where $i=1,\ldots,M$ runs over the set of training data (mesons), and each $x_i$ is a $D$-dimensional input vector, such as $\vec{v}$ in~\eref{eq:vec}. The \textit{covariance} function defines the prior on noisy observations:
\be\label{eq:covfunc}
\text{cov}(x_i, x_j)=k(x_i, x_j)+N_{i,j}~,~~~\text{where~~}N_{i,j}=\sigma_n^2 \delta_{i, j} ~.
\ee
Each positive definite covariance function lends itself to an expansion in terms of basis functions. Two of the simplest and most commonly used \emph{kernels} are the \emph{squared exponential (SE)} and the \emph{rational quadratic (RQ) kernels} defined below,
\begin{align}\label{eq:covfuncSERQ}
k_\text{SE}(x_i, x_j)~&=~\sigma_f^2~\text{exp}\left(-\frac{1}{2}(x_i-x_j)^\text{T} \Lambda^{-1}~(x_i-x_j)~\right)~~~,\nn\\
k_\text{RQ}(x_i, x_j)~&=~~\sigma_f^2~\left(1+\frac{1}{2\alpha}(x_i-x_j)^\text{T} \Lambda^{-1}~(x_i-x_j)\right)^{-\alpha},
\end{align}
where $\Lambda$ is a diagonal matrix with entries $\{\lambda_i^{2}\}_{i=1}^{D}$, where $\lambda_i$ is the characteristic length scale for the $i^\text{th}$ feature, and $\sigma_f$ is an overall scale. We also have ${\alpha>0}$. Since the $D$ distinct characteristic length scales determine the relevance of each input feature, the kernels above implement automatic relevance determination (ARD) in our experiments. The SE covariance function can be expanded in terms of an infinite number of Gaussian basis functions. The RQ covariance function is an infinite sum of SE covariance functions with distinct characteristic length scales and converges to the SE kernel in the limit $\alpha\rightarrow \infty$.  Covariance functions can be combined in various ways yielding complex models. 

Model selection in Gaussian processes is done by maximising the \emph{log marginal likelihood} with respect to a training set $\{x_i\rightarrow y_i\}_{i=1}^M$ of size $M$. This is defined as
\be{\label{eq:Gaussian processLik}}
\log p(y|X) := -\frac{1}{2} y^{T} (K+ N)^{-1} y -\frac{1}{2}\log \left| K+N \right|  -\frac{M}{2} \log (2 \pi ) ~,
\ee
where $y$ is the $M$-vector of training set output values and $X$ is the collection of input vectors of size $D\times M$.

Maximising the \emph{log marginal likelihood} \eqref{eq:Gaussian processLik} sets the hyperparameters of the covariance function \eqref{eq:covfunc}. At the optimal value of the hyperparameters, a trade off is achieved between the complexity of the model and the model fit. Once the hyperparameters are set, one can predict the mean ($\mu_{\star}$) and the variance ($\text{var}_{\star}$) of the distribution for an unseen input vector $x_\star$, using the formulae
\be{\label{eq:Gaussian processpreds}}
\mu_{\star}=k_{\star}^{T}(K+ N)^{-1} y ~, \qquad
\text{var}_{\star}=k(x_\star,x_\star)-k_\star ^{T} (K+ N)^{-1} k_\star ~,
\ee
where the components of the $M$-vector $k_{\star}$ are given by $k_{\star,i} = k( x_\star,x_i)$. 

\subsection{Experimental set-up}\label{exptsetup}
We use the paradigm of supervised machine learning. In our experiments, we use meson data for training and validation. We use a non-overlapping set of all {baryons as the test set}. We also test on exotic hadrons. In the following, we describe how we prepare the datasets before running experiments on them. Following this, we describe the experimental set-up for the neural network and Gaussian process. The results of these experiments are discussed in Section~\ref{sec:four} and summarised in Figures \ref{fig:MesonT}--\ref{fig:bary} and Tables~\ref{tab:tetraquarks1}--\ref{tab:tetraquarks5}.

\paragraph{Encoding particle features:}
There are $109$ unique vectors corresponding to the $196$ mesons in the dataset. Thus, the input vectors do not uniquely identify a particle. For instance, the 13-vector $\vec{v}_\eta = (\vec{0}_4, 1, 1, \vec{0}_6, -1)$, can refer to either $\eta^{\prime}(958)$ or $\eta(1475)$. On average, particles referred to by the same $13$-vector differ in mass by $\sim6$\% (in logarithmic scale) compared to the mean value. Our experiments showed that resolving such \textit{ambiguous inputs} vastly improved the log marginal likelihood for the Gaussian process, but had an adverse effect on the cross-validation error for our neural network. As such, in our experiments involving the neural network, we continue to work with the particle encoding defined in~\eqref{eq:vec}. \comment{reproduced below. 
\begin{equation}\label{NNfeatures}
\text{Train (test) input features (NN):~~~} \vec{v}_\text{tr (test)} ~~= (d,\bar{d},u,\bar{u},s,\bar{s},c,\bar{c},b,\bar{b},I,J,P)~.
\end{equation}}
The inherent ambiguity in the meson dataset may mitigate overfitting of the neural network.

For the Gaussian process, we add an extra input feature\footnote{We also performed the exercise of adding random noise to certain entries of the input 13-vector~\eqref{eq:vec} to resolve ambiguous inputs. However, this did not improve the log marginal likelihood by much when compared to adding an extra feature. Additionally, adding noise to accurately measured quantum numbers is particularly unphysical as the entries in $\vec{v}$ are integers and half-integers.} to distinguish between mesons with the same input $13$-vector $\vec{v}$ in~\eqref{eq:vec}. However, we found no independent physical quantity (that has been measured accurately for all the $196$ mesons) which would serve this purpose. As such, we order these mesons by their masses and use their rank (denoted by \textit{rk}) as the distinguishing feature for the ambiguous inputs.\footnote{We found this to be highly effective in our regression problem, wherein there is a one to many relationship between inputs and outputs in the training data. We posit this as a potentially useful strategy in other regression problems.} The mesons~$\eta^{\prime}(958)$ and $\eta(1475)$ are then encoded by the $14$-vectors $ (\vec{0}_4, 1, 1, \vec{0}_6, -1,0)$ and $ (\vec{0}_4, 1, 1, \vec{0}_6, -1,1)$, respectively. Meson inputs that correspond to a unique mass in the dataset are appended by zero. We append zero to all test inputs. In summary, we then have:
\begin{align}
\text{Train input features (GP):~~~}& \vec{v}_\text{tr} ~~= (d,\bar{d},u,\bar{u},s,\bar{s},c,\bar{c},b,\bar{b},I,J,P, \textit{rk}) ~,\nn\\ 
\text{Test input features (GP):~~~}& \vec{v}_\text{test} = (d,\bar{d},u,\bar{u},s,\bar{s},c,\bar{c},b,\bar{b},I,J,P,\  0 ) ~.\label{eq:vec2}
\end{align}

\paragraph{Encoding particle masses:}
In our experiments involving the neural network, we choose to work with the logarithm of meson masses ($m_s$), since this constrains the range of possible outputs. For the Gaussian process, we further normalise the log masses to zero mean and unit variance.
Since $m_s$ is dimensionful, we effectively divide by $1$ MeV before taking the logarithm.
A more natural reference point might be to define masses in units of $\Lambda_\text{QCD}$ or the mass of the lightest meson, $m_{\pi^0}$.
These normalisations, however, yield larger numerical errors near zero.
Noting that any choice of scale produces a linear offset and will not affect the physics, we deliberately chose one that yields better cross-validation error for the neural network and marginal likelihood for the Gaussian process. 
In order to report errors in the natural (MeV) scale having carried out regression using log masses, we convert our posterior normal distributions to log-normal distributions. If our posterior normal distribution has a mean $\mu$ and standard deviation $\sigma$, the corresponding mean in the natural (MeV) scale is $e^{\mu+\sigma^2/2}$ and the variance is $(e^{\sigma^2}-1)e^{2\mu+\sigma^2}$.

\paragraph{Neural network:} 
First, we perform a five-fold cross-validation for model selection. We restrict ourselves to a single layer network with a view to finding the simplest  model of hadronic masses. This choice is also guided by the relative smallness of our training set. Further, we find that adding extra depth to our network does not improve cross-validation errors. As such, we find the optimal number of nodes to be $50$, the activation function be the logistic sigmoid, and the best optimisation method to be ADAM \cttn{kingma2014adam}. With these fixed, we perform the following two experiments. 

\noindent\textit{Experiment~1:} We randomly select $80\%$ of the meson data to train our network and apply the trained network to the remaining $20\%$ of mesons as well as the entire set of baryons. Note that the mesonic set was used for selecting the network architecture as part of the cross validation; however, the baryonic set was excluded from the network training procedure, and the network never saw it. We obtain predictions on the test set by repeating our experiment $10^3$ times. In each such run, we randomly select $80\%$ of the meson data to train. We also randomly initialise the neural network parameters for each run to remove randomness effects from \textit{e.g.} network initialisation or optimisation. The result of this experiment is discussed in Section~\ref{sec:four} and summarised in Figure \ref{fig:bary}a.
 
\noindent\textit{Experiment~2:} We use the full meson data to train our network and then test on the baryons and exotics. Again, we predict on the test set by repeating our experiment $10^3$ times, with the neural network parameters randomly initialised in each run. The result of this experiment is discussed in Section~\ref{sec:four} and summarised in Figure \ref{fig:bary}b.

These experiments yield simple neural network models of baryon masses. Training errors are indicative of how well the neural network has learned the training data. Therefore, we should be cautious in using these trained neural networks as models for meson masses since  $q\bar{q}$ states (mesons) constitute the training data. Noting this caveat, we report how well the trained neural networks have learned mesonic data. We present this result in Figures \ref{fig:MesonT}a--b. In our particular physical application, there may exist a formula describing baryonic and mesonic masses simultaneously. Our trained networks point to the existence of such a formula. 

These experiments were run in \texttt{Mathematica 12.0.0.0}\cttn{wolfram}. On a laptop, each training round takes approximately $40$s.

\paragraph{Gaussian process:} 
We begin by choosing the covariance function\footnote{We found that combining different kernels such as the SE, RQ, and Mat\'ern kernels returned slightly better marginal likelihoods. However, in keeping with the spirit of building a simple and interpretable machine learning model of hadronic masses, we employ the kernel quoted in~\eqref{eq:covfuncregression}.} below
\be\label{eq:covfuncregression}
\text{cov}(x_i, x_j)~=~k_\text{RQ}(x_i, x_j)~+~\sigma_n^2 \delta_{i, j} ,
\ee
where $k_\text{RQ}$ is as defined in~\eqref{eq:covfuncSERQ} and $x_i$ is a $14$-vector and $i,j =1, \ldots, M$ runs over the $196$ mesons. This choice of the covariance function is ubiquitous in regression problems of our kind and is equivalent to searching over the space of all functions that vary smoothly across different length scales. In addition~\eqref{eq:covfuncregression} contains only $D+2=16$ hyperparameters, yielding a  rather simple Gaussian process model of hadrons. We maximise the log marginal likelihood~\eqref{eq:Gaussian processLik} with respect to the full meson data using {augmented} inputs $\vec{v}_\text{tr}$ as defined in~\eqref{eq:vec2}. We then test the Gaussian process on baryons represented by the {projected} inputs $\vec{v}_\text{test}$. We  discuss these results in Section~\ref{sec:four} and summarise them in Figure \ref{fig:bary}c.

On the training set of mesons encoded by $\vec{v}_\text{tr}$, the Gaussian process is expected to have almost zero training error and variance. The mesons could also be encoded by $\vec{v}_\text{test}$ in~\eqref{eq:vec2}. One might wonder how the Gaussian process trained on mesons encoded by $\vec{v}_\text{tr}$, evaluates on mesons encoded by $\vec{v}_\text{test}$. Effectively, this amounts to testing on mesonic data, albeit, without prior knowledge of the hierarchy in masses (for those mesons indistinguishable by quark composition and $I,J,P$ quantum numbers). The hierarchy is information we utilise during training using meson data encoded by $\vec{v}_\text{tr}$, but not while evaluating the trained Gaussian process on meson data encoded by $\vec{v}_\text{test}$. In this sense, the Gaussian process could be viewed as a suggestive model for the meson masses. This is summarised in Figure \ref{fig:MesonT}c.

We implement the Gaussian process using the package \texttt{GPML 4.2}\cttn{gpml42} in \texttt{Matlab R2019\_b}\cttn{matlab}, which allows for covariance functions with ARD distance measures. 
The training and inference takes $\sim 5$s on a laptop computer.

\subsec{Results}\label{sec:four}

Before invoking a machine learning methodology, an even simpler option to consider is linear regression on the mesonic data. 
As mesonic and baryonic masses spread over roughly four orders of magnitude, we perform the regression over the logarithms of the masses instead of the masses themselves and utilise a $\chi$-squared minimisation. 
While this regression analysis is relatively successful in learning the mesonic sector with $3.2\%$ error (in log scale), the masses are predicted with $42\%$ error (in log scale) in the baryonic sector.
In the expression for mass obtained from regression, the contributions of parity and isospin are essentially negligible.
Linear regression also makes an unphysical distinction between quarks and antiquarks by inferring vastly different masses for them. 

We now describe the results of our non-linear regression using neural networks and Gaussian processes.\footnote{The results for neural networks were presented in\cttn{dkmp,vj}.}

In Figure~\ref{fig:MesonT}, we show how well our machine learning models have learned the meson masses. We include $1\sigma$ error bars computed across $10^3$ runs of the neural network.\footnote{Mass predictions across $10^3$ runs of the neural network fit a normal distribution. We therefore compute standard deviations from the mean to quantify the error in our predictions.} We show the same for the Gaussian process which automatically provides standard deviation from its posterior. Machine learning techniques do not perform well on the pions, which are pseudo-Goldstone bosons corresponding to the spontaneous breaking of chiral symmetry.
The other particles with large standard deviations in Figure~\ref{fig:MesonT}a are the members of the $\pi(1300)$ and $\pi(1800)$ multiplets.
For the data plotted in Figure~\ref{fig:MesonT}b, the mean absolute training error normalised by mass is $0.185$, and the corresponding standard deviation over all the runs is $0.480$.
Since they are much lighter than the other mesons, if we neglect the pions, the absolute errors become $0.130\pm0.170$.
Again neglecting the pions, the Gaussian process predictions yield $0.135\pm0.226$ absolute error.
We emphasise that these values are calculated in terms of the actual masses rather than their logarithms. 

\newcommand\figmesons{
\begin{figure}[H]
\centering
\hspace{30pt}\vspace{-20pt}(a)\\
\includegraphics[scale=.45]{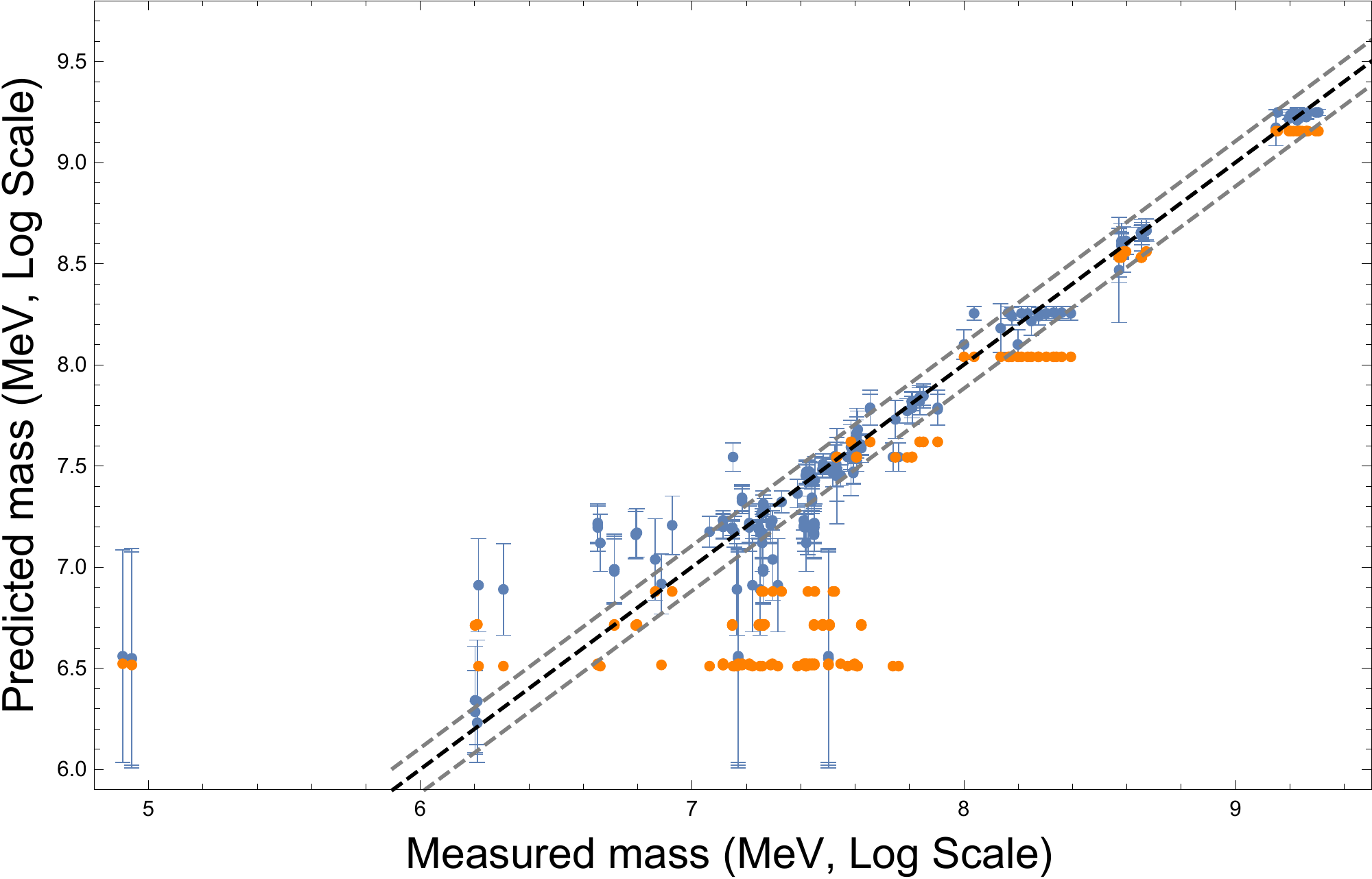}\\[15pt]
\hspace{30pt}\vspace{-20pt}(b)\\
\includegraphics[scale=.45]{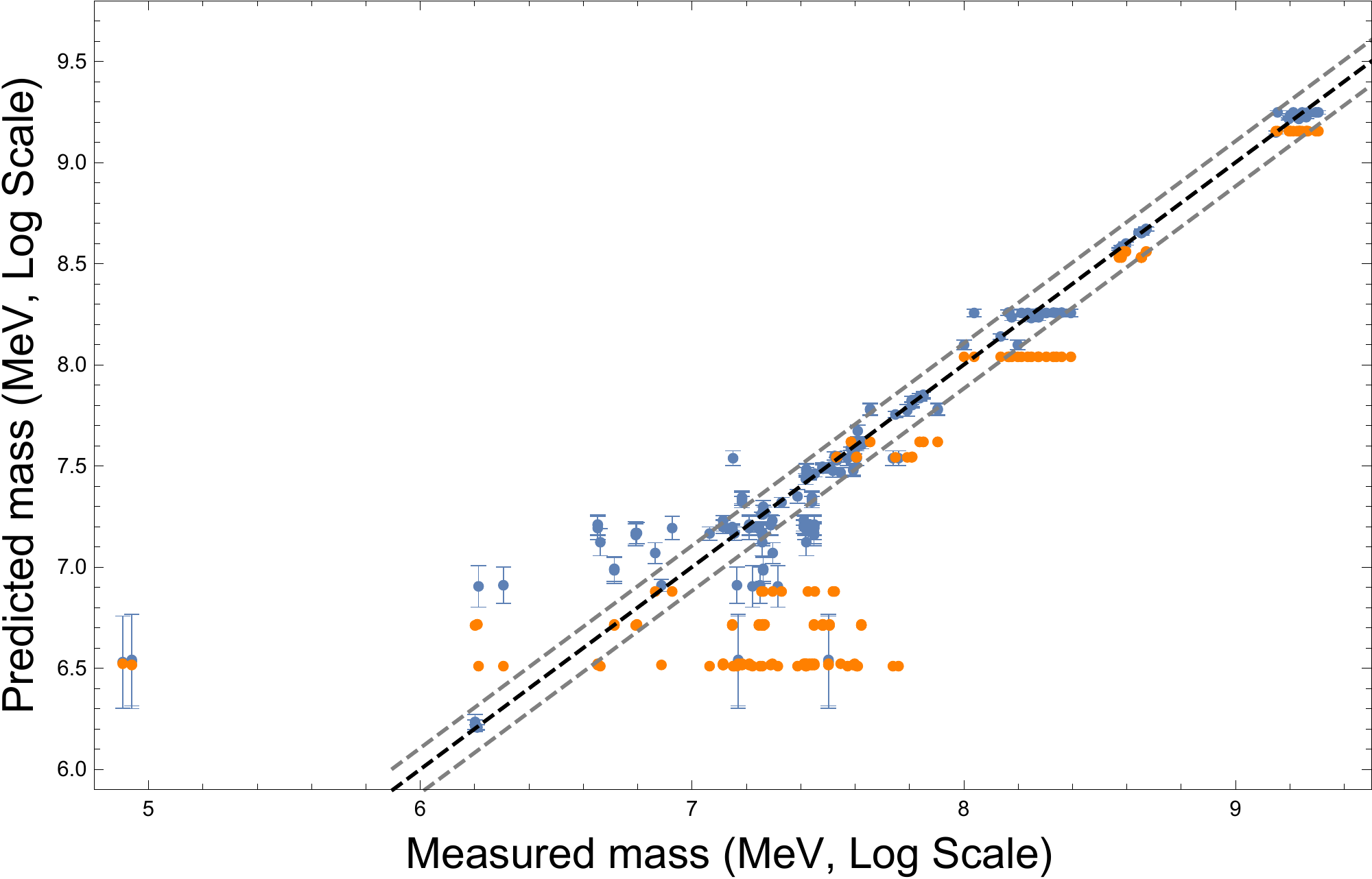}\\[15pt]
\hspace{30pt}\vspace{-20pt}(c)\\
\includegraphics[scale=.45 ]{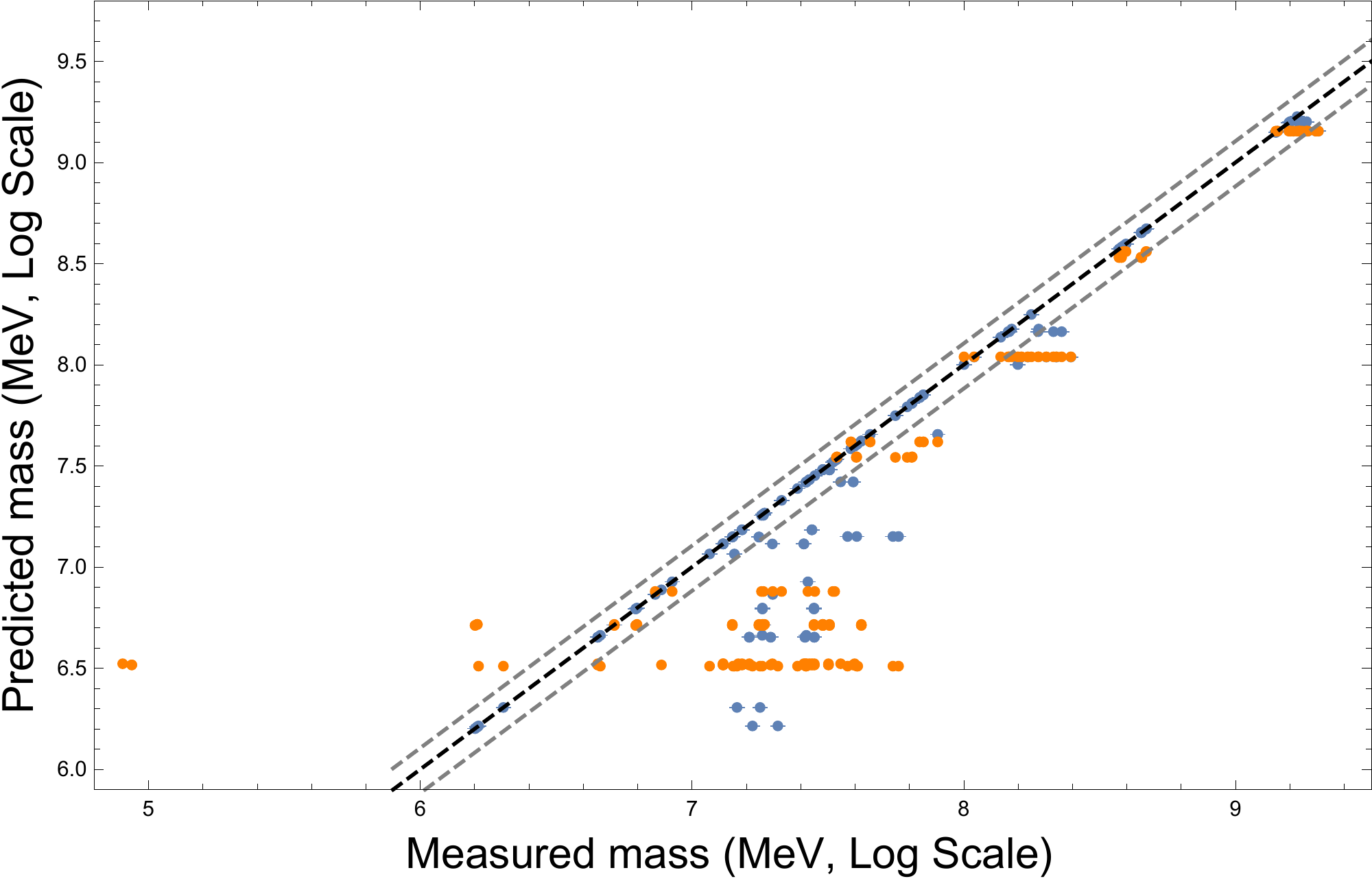}
\caption{\small{\textit{
Learned meson masses upon neural network training with (a) $80$\% and (b) $100$\% of the mesonic data.
In (c) we show the same for the Gaussian process. The dashed lines correspond to $10$\% error in linear mass scale, and the orange dots correspond to the expected mass values from the constituent quark model. Pions are on the far left. The mean training errors are ${18.5\%}$ for the neural network (when seeing the full data) and ${13.3\%}$ for the Gaussian process. 
}}}
\label{fig:MesonT}
\vspace{-10pt}
\end{figure}
}

\newcommand\figbaryons{
\begin{figure}[H]
\centering
\hspace{30pt}\vspace{-20pt}(a)\\
\includegraphics[scale=.45]{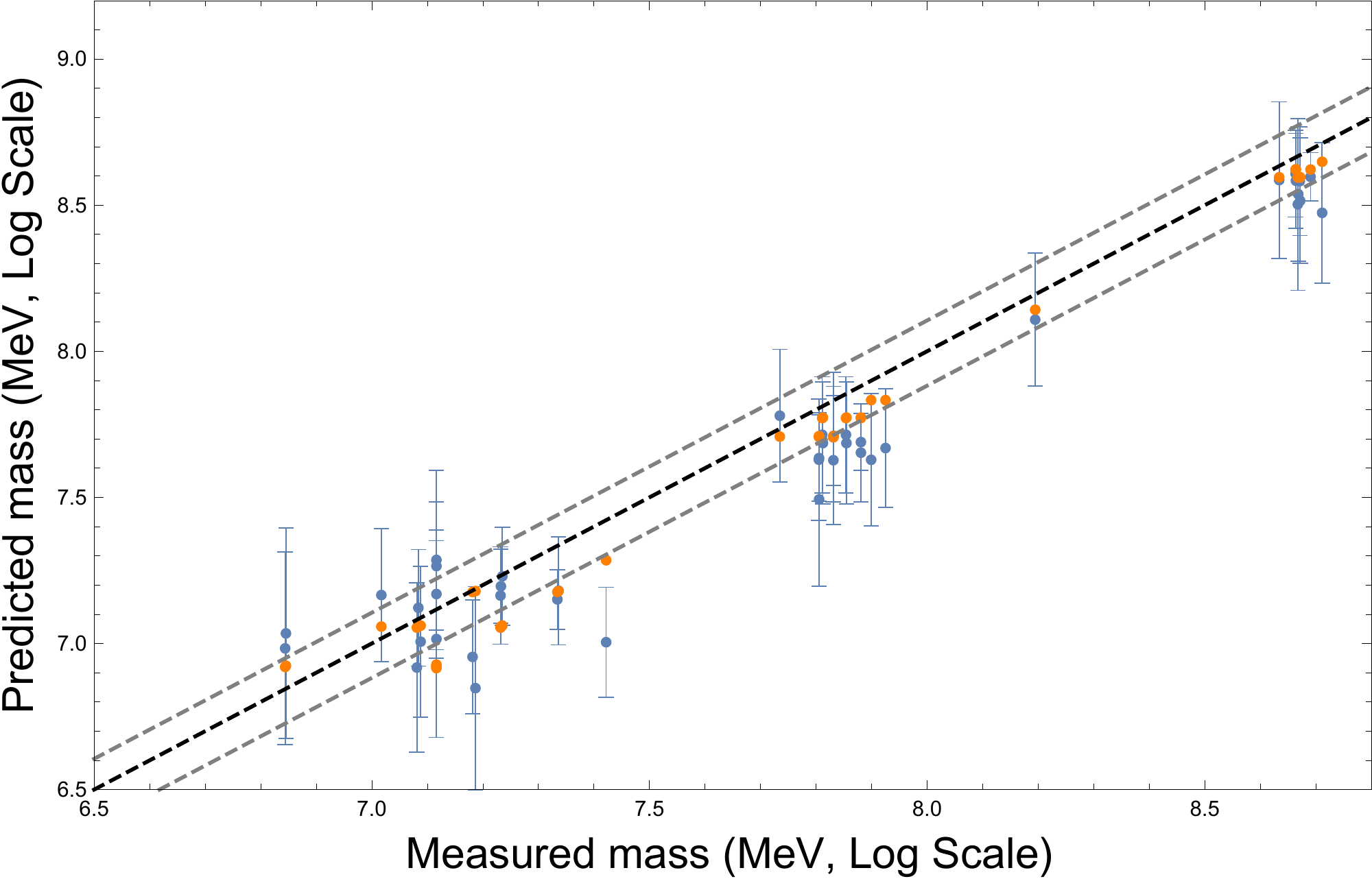}\\[15pt]
\hspace{30pt}\vspace{-20pt}(b)\\
\includegraphics[scale=.45]{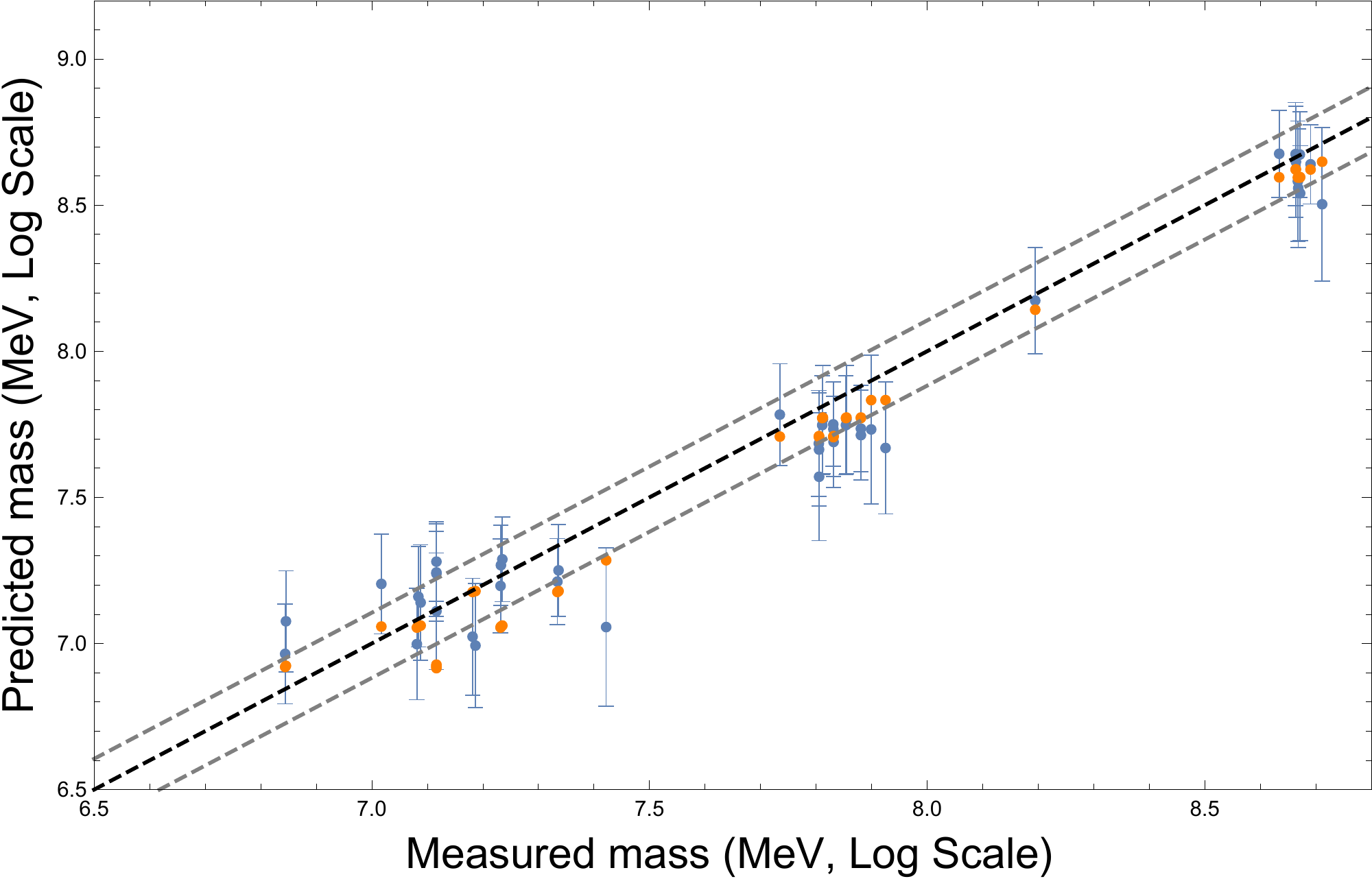}\\[15pt]
\hspace{30pt}\vspace{-20pt}(c)\\
\includegraphics[scale=.45 ]{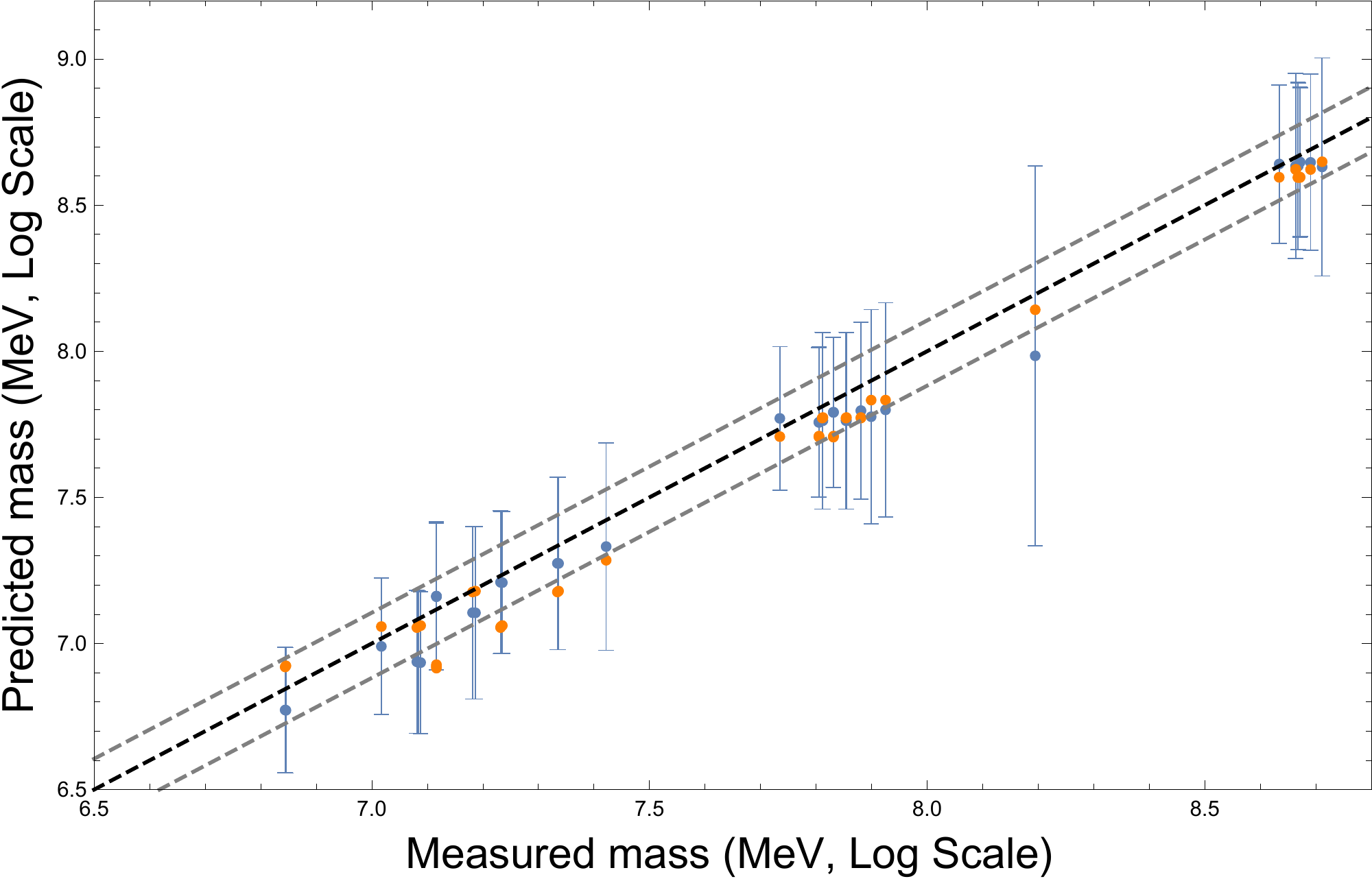}
\caption{\small{\textit{Predicted baryon masses upon neural network training with (a) $80$\% and (b) $100$\% of the mesonic data. In (c) we show the Gaussian process predictions. The dashed lines correspond to $10$\% error in linear mass scale. The results are compared to the predictions of the constituent quark model (in orange). The mean test errors on these predictions are $9.7\%$ for the neural network (trained with the full mesonic data) and ${3.4\%}$ for the Gaussian process. In comparison, the baseline constituent quark model has a mean error of ${8.7\%}$.}} }
\label{fig:bary}
\vspace{-10pt}
\end{figure}
}

After training, we evaluate how the neural network performs on the baryons (the test set).
We consider two cases, with $80$\% and $100$\% of the mesons as training data.
The predictions of the neural network compared to the actual masses are plotted in Figures~\ref{fig:bary}a and~\ref{fig:bary}b.
We observe that in both cases the predictions match the experimental data with remarkable accuracy with the main difference being that the variance widths reduce roughly by a factor of ten if one considers $100$\% instead of $80$\% of the data.

We notice the clustering of masses into three groups.
The first corresponds to baryons composed of light quarks, $u$, $d$, or $s$.
The second group corresponds to baryons containing a $c$ quark, while the third group corresponds to baryons containing $b$ quarks. 
For the test set (baryons), we compute absolute test errors of $0.189\pm 0.154$ (using $80$\% of mesons for training), and $0.097\pm 0.074$ (using $100$\% of mesons for training).
The Gaussian process predictions, shown in Figure~\ref{fig:bary}c, yield an absolute test error of $0.034\pm 0.032$ which is again remarkable. It is noteworthy that our simple Gaussian process model~\eqref{eq:covfuncregression} with as few as $16$ parameters generalises this well to the baryonic data. 

Our results should be compared to the constituent quark model\cttn{Karliner:2003sy}, which assigns an effective mass to each of the valence quarks that accounts for contributions from sea quarks and gluons.
The mass of hadronic states is overwhelmingly due to the confining effects of the strong interaction rather than the Dirac masses of constituent valence quarks.
The simplest effective model sums the masses of the constituent quarks.
These masses are model dependent; we use $m_u = 336\ \text{MeV}$, $m_d = 340\ \text{MeV}$, $m_s = 486\ \text{MeV}$, $m_c = 1550\ \text{MeV}$, $m_b = 4730\ \text{MeV}$.
The predictions of the constituent quark model are shown in orange in Figures~\ref{fig:MesonT} and~\ref{fig:bary}. In the mesonic sector, the constituent quark model has an error of $39.3$\%.
The mean percentage error on the baryon sector is $9.7$\% for the neural network and $3.4$\% for the Gaussian process.
The baseline constituent quark model performs marginally better than the neural network with a mean percentage error of $8.6\%$. 

\figmesons

\figbaryons

It is illuminating to examine the predictions of the neural network on the lightest baryons.
After $10^3$ runs, we consider the ordering in mass for the seven lightest baryons.
The results are presented in Table~\ref{tab:probabilities}.
\begin{table}[h]
\begin{minipage}[b]{0.5\linewidth}\centering
\renewcommand{\arraystretch}{1.4}
{\tiny
\begin{tabular}{|c|c|c|c|c|c|c|c|}\cline{2-8}
\multicolumn{1}{c|}{} & $p$	& $n$	&$\Lambda^0$	& $\Sigma^+$ & $\Sigma^0$ & $\Sigma^-$ & $\Delta^{++}$ \\ \hline
\#1	& 	\cellcolor{green}  22.4 &	10.7	& 1.3  &	11.0	&  0.0    & 4.9     &	7.1 \\ \hline
\#2	& 17.1 &		\cellcolor{green}  13.2	& 2.3	 & 11.1	& 1.6     &	7.2    &	7.6 \\ \hline
\#3	& 15.8 &	12.6	& 	\cellcolor{green}  5.1	 &  7.8	& 3.1	    &  7.3	&  6.7 \\ \hline
\#4	& 10.2 &	10.8	& 6.3	 & 	\cellcolor{green}  8.6	& 4.2	    &  8.3	&  6.9 \\ \hline
\#5	& 8.0	   &   10.3	& 8.3	 & 8.2	& 	\cellcolor{green}  4.4	    &  9.8    & 	5.9 \\ \hline
\#6	& 5.3   &	7.7	& 8.3	 & 6.5	& 7.0     &	\cellcolor{green}  	9.0	&  6.1 \\ \hline
\#7	& 4.9   &	4.3	& 9.6	 & 6.9	& 7.2	     & 7.9    &  	\cellcolor{green}  5.6 \\ \hline
\end{tabular}
}
\end{minipage}
\begin{minipage}[b]{0.5\linewidth}\centering
\renewcommand{\arraystretch}{1.4}
{\tiny
\begin{tabular}{|c|c|c|c|c|c|c|c|}\cline{2-8}
\multicolumn{1}{c|}{} & $p$	& $n$	&$\Lambda^0$	& $\Sigma^+$ & $\Sigma^0$ & $\Sigma^-$ & $\Delta^{++}$ \\ \hline
\#1 &	\cellcolor{green} 79.0   & 	2.0      &	0.0	  &  4.6   &	 0.0   & 2.5  &	5.3 \\ \hline
\#2	 & 13.9 &		\cellcolor{green} 26.7	&  0.0   &	 12.5  &	 1.2 & 5.5  & 22.5 \\ \hline
\#3	 & 4.5   &	21.7  & 	\cellcolor{green} 	0.3 &  15.9	 &  2.2 &  8.6 & 14.7 \\ \hline
\#4	 & 1.9   &   14.7	&  0.6 &  	\cellcolor{green} 17.3 & 	 5.7 &  8.5 &	13.6 \\ \hline
\#5	 & 0.6   &   11.1	&  2.3 &	  15.6 & 	\cellcolor{green}	 6.2 & 11.8 &	9.2 \\ \hline
\#6	 & 0.1   &	9.1	&  3.9 &   9.6	 &  8.0	   & 	\cellcolor{green} 10.7 & 5.8 \\ \hline
\#7	 & 0.0	   &   4.8	&  6.7 &   7.9	 &  11.0  &  9.8	 & 	\cellcolor{green} 5.6 \\ \hline
\end{tabular}
}
\end{minipage}
\caption{\textit{Probabilities for obtaining the correct ordering in ascending mass for the seven lightest baryons in the spectrum.
The left side corresponds to the results for network training with $80$\% of the mesonic data.
The right table corresponds to the results after training with all mesons.
(The numbers in a row or a column do not always sum to $100$ because other outcomes sometimes occur.)}} 
\label{tab:probabilities}
\end{table} 

We observe that when training with $80$\% of the mesonic data, the proton has the highest probability of being the lightest particle. When we use the full meson dataset for training, this probability increases from $22.4$\% to $79.0$\%.
The neutron has the highest probability of being the second lightest particle, but this probability is only $26.7$\% upon training with the full mesonic spectrum.
For the remaining particles, the probabilities are in tension with observation.
For instance, it appears very unlikely for the neural network to predict $\Lambda^0$ as the third lightest particle. 
Using all mesons for training, the neural network predicts
\be
\text{NN:} \quad m_p = 1068\pm 183\ \text{MeV} ~,  \qquad m_n = 1205\pm 206\ \text{MeV} ~, \label{eq:mpmn}
\ee
which are accurate to $13.9$\% and $28.2$\% of the measured values.\footnote{
The measured masses of the proton and neutron are $m_p = 938.28$ MeV and $m_n = 939.57$ MeV.}
While the ordering is correct, the mass difference is significantly larger than we expect.
The Gaussian process, meanwhile, yields superior predictions: 
\be
\text{GP:} \quad m_p = 893.8\pm194.4\ \text{MeV} ~, \qquad  m_n = 892.8\pm 193.9\ \text{MeV} ~; \label{eq:mpmn}
\ee
these are within $4.7$\% and $5.0$\% of the measured values. 
Noting the error bars, the Gaussian process predicts the proton and neutron to be almost degenerate.

\subsection{Predictions for other resonances}\label{sec:four.1}
We would like to evaluate the performance of our machine learning algorithms on resonances beyond the meson and baryon spectrum, \textit{i.e.}, resonances that do not fit in the $q\bar{q}$ or the $qqq/\bar{q}\bar{q}\bar{q}$ pictures.
Additional states could correspond to glueballs, hybrid mesons (that can be viewed as a $q\bar{q}$ state with a gluon flux tube in such a way that the gluon contributes to the global $J^{PC}$ quantum numbers of the state), as well as combinations with a higher number of quarks such as $qq\bar{q}\bar{q}$, which could correspond either to a compact tetraquark state or to a mesonic molecule, pentaquarks $qqqq\bar{q}$ (either molecular or compact), and baryonium resonances of a baryon and an antibaryon $qqq\bar{q}\bar{q}\bar{q}$.
All of these possibilities are allowed by the representation theory of $SU(3)$. 

The nature of the observed resonances is still a matter of debate.
In many cases, the effective theoretical models do not permit us to disentangle the constituents of a given particle as certain experimental features such as widths, masses, and decay rates cannot be completely reproduced by any of the models available.
Also, since a resonance could result from non-trivial mixings, it might be that the features of a given experimental peak could have simultaneous contributions from, say, a tetraquark and a hybrid meson. 

Since in our case the input variables are taken to be global features of a resonance, we consider only particles that are made out of quarks such as the tetraquarks, penataquarks, baryon--antibaryon states, and dibaryons (such as $^2{\rm H}^+$ and $^2{\rm He}^{++}$).
Also, since we cannot specify whether a particle is compact or molecular, we simply give the quark content as an input along with $I,J,P$ quantum numbers.
We focus on the states most widely studied which are summarised in reviews in the PDG\cttn{PhysRevD.98.030001}.
For likely compositions, we refer to the reviews\cttn{Chen:2016qju,Liu:2019zoy}, which summarise the theoretical attempts to model pentaquarks and tetraquarks. 
In particular, for the tetraquarks we focus on three different types of systems which fit relatively well into the molecular picture.
Neural network results are computed from $10^3$ runs of training and testing.
The Gaussian process sometimes predicts the same masses when confronted with different possibilities for the composition of a given resonance.
Comparing this observation to the cartoon of the Gaussian process from Figure~\ref{fig:GPtoy}, this indicates that the moduli space around certain points is narrow, and there is effectively a flat direction. 
Unreasonably small errors in some of the Gaussian process predictions, for example in Table~\ref{tab:tetraquarks2} and Table~\ref{tab:tetraquarks5}, correspond to a test vector of mesonic type being coincident with a vector used for training.

\paragraph{Light--Light systems:} 
Here, we concentrate on the $a_0(980)$ and $f_0(980)$, which lie slightly below the $K\overline{K}$ threshold.
Their composition still remains an open question:
they could be conventional mesons, tetraquark states, $K\overline{K}$ molecules, or a meson--meson system mixed with a scalar glueball.
Here, we test how the neural network and the Gaussian process perform on two hypotheses: conventional meson vs.\ tetraquark. 
The results of this experiment are given in Table~\ref{tab:tetraquarks1}.
\begin{table}[h]
\begin{center}
\renewcommand{\arraystretch}{1.4}
{\tiny
\begin{tabular}{|c|c|c|c|c|c|c|}\cline{2-7}
\multicolumn{1}{c|}{} &  $I\,(J^P)$	 &  Measured mass (MeV) &  Composition & NN Pred (MeV) & GP Pred (MeV) & CQM Pred (MeV) \\ \hline
\multirow{2}{*}{$a_0(980)$}	 &\multirow{2}{*}{ $1\,(0^+)$}	& \multirow{2}{*}{$980\pm 20$} & $u\bar{u}$    &   $1277\pm 246$ &   $511\pm {34}$   & $680$\\ \cline{4-7}
& & & $u\bar{u}s\bar{s}$ $(K\overline{K})$   &   $2172\pm 466$  & $1713\pm {68}$   & $1652$ \\ \hline
\multirow{2}{*}{$f_0(980)$}	 &\multirow{2}{*}{ $0\,(0^+)$}	& \multirow{2}{*}{$990\pm 20$} & $d\bar{d}$   &   $921\pm 117$ &    $977\pm {37}$  & $672$\\ \cline{4-7}
& & & $d\bar{d}s\bar{s}$ $(K\overline{K})$   &   $1592\pm 401$   &  $1132\pm {312}$ & $1644$\\ \hline
\end{tabular}
}
\caption{\label{tab:tetraquarks1} \small{\textit{Machine learning and baseline constituent quark model predictions for light--light systems.
}} }
\end{center}
\end{table} 

\paragraph{Heavy--Light systems:}
In the heavy--light sector, we consider the $D^*_{s0}(2317)^\pm$ as well as the $D_{s1}(2460)$, which have been interpreted as tetraquarks or $DK$ ($D^*K$) molecules. 
The results of this experiment are given in Table~\ref{tab:tetraquarks2}.
\begin{table}[h]
\begin{center}
\renewcommand{\arraystretch}{1.4}
{\tiny
\begin{tabular}{|c|c|c|c|c|c|c|}\cline{2-7}
\multicolumn{1}{c|}{} &  $I\,(J^P)$	 &  Measured mass (MeV) &  Composition & NN Pred (MeV) & GP Pred (MeV)  & CQM Pred (MeV) \\ \hline
\multirow{2}{*}{$D_{s0}^*(2317)^\pm$}	 &\multirow{2}{*}{ $0\,(0^+)$}	& \multirow{2}{*}{$2317.8\pm0.5$} & $c\bar{s}$   &   $2640\pm 433$  & $2434\pm {700}$ & $2036$ \\ \cline{4-6}\cline{7-7}
& & & $c\bar{u}u\bar{s}$ $(DK)$   &   $4326\pm 925$   & $2474\pm {826}$ & $2858$\\ \hline
\multirow{2}{*}{$D_{s1}(2460)^\pm$}	 &\multirow{2}{*}{ $0\,(1^+)$}	& \multirow{2}{*}{$2459.5\pm 0.6$} & $c\bar{s}$    &   $2547\pm 39$   & $2535\pm {0.003}$ & $2036$\\ \cline{4-6}\cline{7-7}
& & & $c\bar{u}u\bar{s}$ $(D^*K)$   &   $3431\pm 544$  &  $2560\pm {788}$  & $2858$\\ \hline
\end{tabular}
}
\caption{\label{tab:tetraquarks2} \small{\textit{Machine learning and baseline constituent quark model predictions for heavy--light systems.
}}}
\end{center}
\end{table} 

\paragraph{Heavy--Heavy systems:} 
We consider some of the states of the charmonium sector: $X(3872)$, $Y(4260)$, $Y(4360)$, and $Y(4460)$ as well as the charged ones $Z_c(3900)^\pm$, $Z_c(4020)^\pm$, $Z_c(4200)^\pm$ and $Z_c(4430)^\pm$.
Similarly, for the bottomonium sector, we have $Z_b(10650)^\pm$. 
The results of this experiment are given in Table~\ref{tab:tetraquarks3}.
\begin{table}[h]
\begin{center}
\renewcommand{\arraystretch}{1.4}
{\tiny
\begin{tabular}{|c|c|c|c|c|c|c|}\cline{2-7}
\multicolumn{1}{c|}{} & $I\,(J^P)$	 & Measured mass (MeV) & Composition	&  NN Pred (MeV) & GP Pred (MeV) & CQM Pred (MeV)\!\!\! \\ \hline
$X(3872)$	 
& $0\,(1^+)$	& $3871.69\pm 0.17$  & $c\bar{u}\bar{c}u$ $(D^0\,\bar{D}^{*0})$ &   $4815\pm 786$ & $3514\pm {190}$  & $3772$\\ \hline
$Y(4260)$ & $ 0\,( 1^-)$ & $4230\pm 8$  & $c\bar{s}s\bar{c}$ $(D_s\,\bar{D}_s)$ &  $(5.4\pm 1.1)\times 10^3$  & $3543\pm {1167}$ & $4072$ \\ \hline
$Y(4360)$ & $ 0\,( 1^-)$&  $4368\pm 13$  &  $c\bar{u}\bar{c}u$ $(D_1\,\bar{D}^*)$ & \multirow{2}{*}{$4940\pm 903$} & \multirow{2}{*}{$3107\pm {168}$} &  \multirow{2}{*}{$3772$}  \\ \cline{1-4}
$Y(4660)$ & $ 0\,( 1^-)$ & $4643\pm 9$  &  $u\bar{u}c\bar{c}$ $(f_0(980)\,\psi^\prime)$ &  &  & \\ \hline
$Z_c(3900)^\pm$ & $ 0\,( 1^+)$	&$3886.6\pm 2.4$  & $c\bar{d}\bar{c}u$ $(D\,\bar{D}^*)$ &  \multirow{4}{*}{$4991\pm 815$} & \multirow{4}{*}{$3515\pm {199}$} & \multirow{4}{*}{$3776$} \\  \cline{1-4}
$Z_c(4020)^\pm$ &$ 0\,( 1^+)$	&$4024.1\pm 1.9$  & $c\bar{d}\bar{c}u$  $(D^*\,\bar{D}^*)$ &  & &\\  \cline{1-4}
$Z_c(4200)\pm$ & $ 0\,( 1^+)$	&  $4196^{+35}_{-32}$  & $cu\bar{c}\bar{d}$ &   &   &\\  \cline{1-4}
$Z_c(4430)^\pm$ &$ 0\,( 1^+)$	&$4478^{+15}_{-18}$ & $c\bar{d}\bar{c}u$  $(D_1 D^*\,,\,D_1^\prime D^*)$   &    &  &\\ \hline 
$Z_b(100610)^\pm$ & $ 0\,( 1^+)$	& $10607.2\pm2.0$  & $b\bar{d}\bar{b}u$  $(B \bar{B}^*)$ &  \multirow{2}{*}{ $(1.47\pm 0.17) \times 10^4$} & \multirow{2}{*}{$9907\pm {560}$}  & \multirow{2}{*}{$10136$} \\  \cline{1-4}
$Z_b(100650)^\pm$ & $ 0\,( 1^+)$	&$10652.2\pm1.5$  &$b\bar{d}\bar{b}u$ $(B^* \bar{B}^*)$ &  & & \\ \hline
\end{tabular}
}
\caption{\label{tab:tetraquarks3} \small{\textit{Machine learning and baseline constituent quark model predictions for heavy--heavy systems. The mean errors in the machine learning predictions are $25.0\%$ (neural network) and ${15.5\%}$  (Gaussian process). In comparison, the  constituent quark model has an error of ${9.1\%}$.}}}
\end{center}
\end{table}

\paragraph{Pentaquarks:}
We consider the three so-called charmonium pentaquarks observed by the LHCb experiment.
These states have been identified as sharp resonances appearing as intermediate states in $\Lambda_b$ baryon decays and have been interpreted either as compact pentaquarks or as baryon--meson molecules.
For our purposes, it suffices to describe their quark composition as $uudc\bar{c}$ and the corresponding quantum numbers and masses are shown in the following table together with our machine learning predictions. The results of this experiment are given in Table~\ref{tab:pentaquarks}.
\begin{table}
\begin{center}
\renewcommand{\arraystretch}{1.4}
{\tiny
\begin{tabular}{|c|c|c|c|c|c|}\cline{2-6}
\multicolumn{1}{c|}{$uudc\bar{c}$} & $I\,(J^P)$ & Measured Mass (MeV) & NN Pred (MeV) & GP Pred (MeV) & CQM Pred (MeV)\\ \hline
${\rm P}_c(4312)^+$ &  $\frac12\,(\frac12^+)$ & $4311.9\pm0.7^{+6.8}_{-0.6}$ & $(4.2\pm 1.2)\times 10^3$ & $3544\pm {923}$ & \multirow{3}{*}{$4112$}  \\ \cline{1-5}
${\rm P}_c(4440)^+$ &  $\frac12\,(\frac12^-)$ &$4440.3\pm1.3^{+4.1}_{-4.7}$ &  $(4.1\pm 1.1)\times 10^3$ & $3253\pm {846}$&\\  \cline{1-5}
${\rm P}_c(4457)^+$ &  $\frac12\,(\frac32^-)$ & $4457.3\pm0.6^{+4.1}_{-1.7}$ &  $(4.5\pm 1.1)\times 10^3$ & $3581\pm {932}$ &\\ \hline
\end{tabular}
}
\caption{\label{tab:pentaquarks}  \small{\textit{Machine learning and baseline constituent quark model predictions for pentaquarks. The mean errors in the machine learning predictions are ${3.7}\%$ (neural network) and $21.3\%$  (Gaussian process). In comparison, the   constituent quark model has an error of ${6.5\%}$}.}}
\end{center}
\end{table}
\paragraph{Baryonium:}
Finally, we consider the $f_2(1565)$ which have been only observed in $p\bar{p}$ annihilation processes and could be a bound $p\bar{p}$ state, together with the dibaryons $^2{\rm H}^+$ and $^2{\rm He}^{++}$. 
The results of this experiment are given in Table~\ref{tab:tetraquarks5}.
\begin{table}[h]
\begin{center}
\renewcommand{\arraystretch}{1.4}
{\tiny
\begin{tabular}{|c|c|c|c|c|c|c|}\cline{2-7}
\multicolumn{1}{c|}{} &  $I\,(J^P)$	 &  Measured mass (MeV) &  Composition & NN Pred (MeV) & GP Pred (MeV) & CQM Pred (MeV)\\ \hline
\multirow{2}{*}{$f_2(1565)$}	 &\multirow{2}{*}{ $0\,(2^+)$}	& \multirow{2}{*}{$1562\pm13$} & $u\bar{u}$   &   $1879\pm 70$   & $1275\pm {0.002}$ & $672$\\ \cline{4-7}
& & & $uu \bar{u}\bar{u}d\bar{d}$ $(p\bar{p})$   &   $2648\pm 613$ &  $1284\pm {85}$ & $2024$ \\ \hline
$^2{\rm H}^+$ & $0\,(1^+)$ &$1875$& $uuuddd$  $(pn)$   &   $1585\pm 551$ &   $1167\pm138$  & $2028$\\ \cline{1-7}
$^2{\rm He}^{++}$ & $0\,(1^+)$ &$1878$& $uuuudd$  $(pp)$   &   $1409\pm {484}$ & $1166\pm163$  &$2024$  \\ \hline
\end{tabular}
}
\caption{\label{tab:tetraquarks5}  \small{\textit{Machine learning and baseline constituent quark model predictions for (would-be) baryon--antibaryon and dibaryon states. 
}}}
\end{center}
\end{table} 

\subsec{Discussion}\label{sec:five}
The labeling scheme we have chosen has an inherent ambiguity.
The light mesons are linear combinations of $q\bar{q}$ pairs:
\be
\left( \ba{c} \ket{\pi^0} \cr \ket{\eta} \cr \ket{\eta'} \ea \right) =
\left( \ba{ccc} -\frac{1}{\sqrt2} & \frac{1}{\sqrt2} & 0 \cr \frac{1}{\sqrt6} & \frac{1}{\sqrt6} & -\frac{2}{\sqrt6} \cr \frac{1}{\sqrt3} & \frac{1}{\sqrt3} & \frac{1}{\sqrt3} \ea \right) \left( \ba{c} \ket{d\bar{d}} \cr \ket{u\bar{u}} \cr \ket{s\bar{s}} \ea \right) ~. \label{eq:basis}
\ee
We are treating $\pi^0$ as if it was simply a $d\bar{d}$ pair. 
If we interpret~\eref{eq:basis} as a system of linear equations that expresses $m_{d\bar{d}}$, $m_{u\bar{u}}$, and $m_{s\bar{s}}$ in terms of the masses of the physical mesons, we discover that $m_{u\bar{u}} > m_{d\bar{d}} > m_{s\bar{s}}$.
This is opposite to the order of the masses of the bare quarks $u$, $d$, and $s$.
Instead, if we square the relation, we can solve for $(m_{u\bar{u}},m_{d\bar{d}},m_{s\bar{s}}) \approx (234\ \text{MeV}, 425\ \text{MeV}, 1000\ \text{MeV})$.
Defining new input vectors corresponding to the $d\bar{d}$, $u\bar{u}$, $s\bar{s}$ and dropping isospin as $\pi^0$, $\eta$, and $\eta'$ have different values of $I$, we find the neural network performs less well than with the Monte Carlo particle numbering scheme of the PDG\cttn{PhysRevD.98.030001}.

Alternatively, we can think of $\ket{\pi^0}$, $\ket{\eta}$, and $\ket{\eta'}$ as orthonormal eigenstates of the mass squared operator, with, \textit{e.g.},
\be
\bra{\pi^0} \hat{M}^2 \ket{\pi^0} = m^2_{\pi^0} ~.
\ee
Rotating bases, certain matrix elements are degenerate:
\be
\bra{d\bar{d}} \hat{M}^2 \ket{d\bar{d}} = \bra{u\bar{u}} \hat{M}^2 \ket{u\bar{u}} ~,
\ee
and again the neural network does not perform well under this assumption.
Moreover, there are cross terms such as $\bra{d\bar{d}} \hat{M}^2 \ket{u\bar{u}}$ that we do not account for.
As such, the results of our analysis are sensitive to how the data is captured. 

Comparing Figures~\ref{fig:MesonT}a and~\ref{fig:MesonT}b, and Figures~\ref{fig:bary}a and~\ref{fig:bary}b, we notice that with the same network architecture, the size of the dataset used for training markedly improves the performance of the neural network.
Moreover, the neural network successfully predicts that the proton is the lightest baryon $79$\% of the time and in a plurality of cases concludes that the neutron is the next lightest particle.
This ordering is striking because, since $\pi^0$, the lightest meson, is effectively encoded as $d\bar{d}$, we would na\"{\i}vely expect that the $udd$ combination would be lighter than the $uud$ combination.
The neural network learns this subtlety.

Disentangling how a neural network evaluates data to make predictions is a notoriously difficult problem in machine learning.
The weight matrix $W^{(1)}$ corresponding to the hidden layer of our neural network is a $50\times 13$ matrix.
When analysing the various entries of the matrix $W^{(1)}$, we observe relatively uniform contributions to the $50$ neurons.
We also note that the highest values are related to the $c$, $\bar{c}$, $b$ and $\bar{b}$ columns.
Isospin has an intermediate contribution while the smallest contributions come from the three lightest quarks/antiquarks as well as parity and angular momentum.
Despite these contributions being small, they determine the light meson/baryon spectrum. 
The eigenvalue spectrum of the Hermitian matrix $(W^{(1)})^T W^{(1)}$ is robust when averaged over multiple runs. 
This means that the neural network treats the entries of the input vector the same way each time it is trained.
To determine how each component of the input vector is weighted, we can of course test the neural network on unit vectors corresponding to a single non-zero entry in $\vec{v}$.
Performing regression on these outputs does not correlate well to the predictions of the fully connected, trained neural network.
We also do not recover the effective particle masses from the constituent quark model.
As a check, we have used the neural network to predict masses of antibaryons.
These disagree from the baryon prediction by at most $6$\% (log scale) of the actual mass. 
\comment{
Additionally in Figure~\ref{fig:anti}, we show the performance of the net both on baryons and antibaryons.
As one can see there is a systematic offset between the two. 
\begin{figure}[h]
\centering
\includegraphics[scale=.3]{BA}
\caption{\label{fig:anti} Predicted masses for the baryon and antibaryon sectors.}
\end{figure}}

The success of the neural network applied to baryons may hint at the existence of a formula $m(\vec{v})$ that computes the mass based on the valence quark composition and quantum numbers.
A useful comparison to make in this regard is to recall the semi-empirical mass formula\cttn{Weizsacker1935}, which predicts the mass of a nucleon as a function of the number of protons and neutrons with coefficients determined by fitting to experimentally measured masses.
Witten's method for estimating the mass of the baryons in the large-$N$ limit relies on the fact that the interaction between quarks can be approximated by means of a Hartree--Fock type of potential\cttn{Witten:1979kh}.
The Hartree method can also be used in atoms with many electrons and in general distributions in which the density grows as the number of particles increases.
In the case of nuclei, the density does not increase with the number of particles precisely because the volume grows roughly as the number of nucleons.
The binding energy becomes negligible as the total number of nucleons and the number of protons increases.
In the case of baryons at large-$N$ or atoms with many electrons, the interaction energy grows with $N$ as well and cannot be neglected.
This highlights a structural difference between nuclei and baryons.
Therefore, from this perspective it is unclear if baryon masses obey a relation as simple as the one derived from the liquid drop model for nuclei.

Suppose, however, that some more complicated association $m(\vec{v})$ exists.
The Universal Approximation Theorem\cttn{Cybenko, Hornik} establishes that a feed-forward neural network using a sigmoid activation function with a single hidden layer and a finite number of neurons can approximate continuous functions on compact subsets of $\mathbb{R}^n$.
We are not, however, guaranteed that the network that approximates this function has a simple architecture.
It is therefore remarkable that in the work that we report here a single hidden layer consisting of only $50$ neurons performs so well.
This is particularly striking in view of the fact that the training is accomplished with only $196$ mesons.
By contrast, most machine learning efforts in mathematics and physics employ significantly larger datasets.
(See References\cttn{He:2017set,Krefl:2017yox,Ruehle:2017mzq,Carifio:2017bov,Bull:2018uow,Jejjala:2019kio} for representative use cases.)

The Gaussian process is the limit of infinite width for a neural network. Both the neural network and the Gaussian process perform quite well on mesons and baryons. Indeed, the Gaussian process predicts the baryon masses with $96.6$\% accuracy. We observe in Figure~\ref{fig:MesonT} that while the mesonic predictions of the neural network with different training sets and the Gaussian process are essentially the same, the variances with the neural networks are significantly smaller. A similar effect can be seen on the baryons.  Looking at the performance on tetraquarks and pentaquarks, we find that the Gaussian process leads to better predictions on tetraquark masses, even though it gives a single value for all the resonances considered in the charmonium sector. The neural network has a superior performance on pentaquarks compared to the Gaussian process. This suggests that in certain regimes, an architecture with a finite number of neurons works better.

With regard to the exotics, we identify some differences between the neural network and the Gaussian process: for the $a_0(980)$, $f_0(980)$, $D^*_{s0}(2317)$, $D_{s1}(2460)$ and $f_2(1565)$ we tested against different composition hypotheses for the same resonance. In the case of the $a_0(980)$ and $f_0(980)$ particles, both the neural network and the Gaussian process give results which are favorable to a mesonic nature for these. In the other cases, while the neural network predicts different masses for each composition with the closest value to the measured mass being the one corresponding to the mesonic composition, the result from the Gaussian process is almost identical for each of the hypotheses, and hence we cannot use the Gaussian process to discriminate between different quark compositions.  
The Gaussian process predicts the proton and the neutron masses to be almost identical.
Therefore, it is not surprising that the same degenerate results occur when the Gaussian process is asked to predict the masses of $^2\text{H}^+$ and $^2\text{He}^{++}$.

The machine learning algorithms employed in this work take the mesonic data and extrapolate to baryons as well as other resonances with an improved accuracy than the one expected from a simple constituent quark model.
One significant sector of QCD that it is out of reach in this work are glueballs.
It would be interesting if we could improve our methods to describe gluonic contributions to the particle masses and in turn be able to make predictions on glueball masses.
Perhaps a more unified approach to describing mesons and glueballs could be to consider the corresponding Regge trajectories and use that information to characterize gluon plus quark contributions to the composition of a given state.
We can as well envision machine learning aspects of the QCD string.
Additionally, it would also be interesting to use our results to match the coefficients of the Gell-Mann--Okubo formula\cttn{GellMann:1961ky,Okubo:1961jc}, in particular if by using the lower mass coefficients for certain multiplets, we could then predict the coefficients for higher resonances in both the meson and baryon sectors.

We have seen that knowledge of the meson spectrum alone is sufficient to approximate masses of both baryons and exotic colour singlet bound states of more than three quarks.
The accuracy of these predictions lends credence to the suggestion that baryons are solitons in a weakly coupled effective theory of mesons with masses that scale like the inverse coupling $(\frac1N)^{-1} = N$\cttn{Witten:1979kh}.
On theoretical grounds, the success of this idea applied to real world QCD relies on $\frac1{N^2} = \frac19$ dressed by the appropriate phase space factors being small enough for perturbation theory to be reliable and in consequence, the approximate decoupling of physical degrees of freedom in an $S$-matrix picture.
As a next step, we aim to achieve a better understanding of how a machine learns aspects of QCD and compare this to traditional methods in quantum field theory.
Perhaps machine learning techniques can lead to an effective theory that improves chiral perturbation theory.
It would also be interesting to apply machine learning techniques to further develop the theoretical structure of amplitudes in QCD.
We leave these challenges for future work.

\section*{Acknowledgements}
YG and CM are supported by the Defence and Security Programme of the Alan Turing Institute. VJ is supported by the South African Research Chairs Initiative of the Department of Science and Technology and the National Research Foundation.
DKMP is supported by the Simons Foundation Mathematical and Physical Sciences Targeted Grants to Institutes, Award ID:509116.
The authors thank Chris Bouchard, Amanda Cooper--Sarkar, Robert de Mello Koch, Amihay Hanany, Koji Hashimoto, Djordje Minic, Brent Nelson, Tim Rudner,  Subir Sarkar,  Lewis Smith, Wati Taylor and Mark van der Wilk for discussions and comments on the manuscript. 

\bibliography{sample}

\bibliographystyle{JHEP}

\end{document}